\documentclass[11pt, draftclsnofoot, journal, onecolumn]{IEEEtran}
\usepackage{blindtext}
\usepackage{amsmath}
\interdisplaylinepenalty=2500\usepackage{amssymb}
\usepackage{bm}
\usepackage{mdwmath}
\usepackage{graphicx}
\usepackage{xcolor}
\usepackage{color}
\usepackage{subfigure}
\usepackage{makecell}
\usepackage{multirow}
\usepackage{acronym}
\usepackage[noadjust]{cite}
\usepackage{url}
\usepackage{algorithm}
\usepackage{algpseudocode}
\usepackage{hyperref}
\hypersetup{%
	bookmarksopen=true,%
	bookmarksnumbered=true,%
	pdfpagemode={UseOutlines},
	pdfstartview={FitH},%
	colorlinks=true,%
	citecolor={cyan}}
\usepackage{enumitem}
\newlist{abbrv}{itemize}{1}
\setlist[abbrv,1]{label=,labelwidth=1in,align=parleft,itemsep=0.1\baselineskip,leftmargin=94pt}

\renewenvironment{IEEEbiography}[1]
{\IEEEbiographynophoto{#1}}
{\endIEEEbiographynophoto}
\begin{document}
	
	\title{Intelligent Network Slicing for V2X Services Towards 5G}
\author{Jie~Mei,~\IEEEmembership{Student Member,~IEEE,}
	Xianbin~Wang,~\IEEEmembership{Fellow,~IEEE}~and
	Kan~Zheng$^{*}$,~\IEEEmembership{Senior Member,~IEEE,}
	\thanks{Jie~Mei is with the Intelligent Computing and Communication ($ \text{IC}^\text{2} $) Lab, Wireless Signal Processing and Network (WSPN) Lab, Key Laboratory of Universal Wireless Communication, Ministry of Education, Beijing University of Posts and Telecommunications (BUPT), Beijing, 100876, China, and also with Electrical and Computer Engineering, Western University, London, ON N6A 5B9, Canada. (E-mail: meijie.wspn@bupt.edu.cn, jmei28@uwo.ca).}%
	\thanks{Xianbin~Wang is with Department of Electrical and Computer Engineering, Western University, London, ON N6A 5B9, Canada
		(E-mail: xianbin.wang@uwo.ca).}
	\thanks{Kan~Zheng is with the the Intelligent Computing and Communication ($ \text{IC}^\text{2} $) Lab, Wireless Signal Processing and Network (WSPN) Lab, Key Laboratory of Universal Wireless Communication, Ministry of Education, Beijing University of Posts and Telecommunications (BUPT), Beijing, 100876, China (E-mail: zkan@bupt.edu.cn).}
	}
\maketitle	
	
	\begin{abstract}
		Benefiting from the widely deployed LTE infrastructures, the fifth generation (5G) wireless networks has been becoming a critical enabler for the emerging vehicle-to-everything (V2X) communications. However, existing LTE networks cannot efficiently support stringent but dynamic requirements of V2X services. One effective solution to overcome this challenge is network slicing, whereby different services could be supported by logically separated networks. To mitigate the increasing complexity of network slicing in 5G, we propose to leverage the recent advancement of Machine Learning (ML) technologies for automated network operation. Specifically, we propose intelligent network slicing architecture for V2X services, where network functions and multi-dimensional network resources are virtualized and assigned to different network slices. In achieving optimized slicing intelligently, several critical techniques, including mobile data collection and ML algorithm design, are discussed to tackle the related challenges. Then, we develop a simulation platform to illustrate the effectiveness of our proposed intelligent network slicing. With the integration of 5G network slicing and ML-enabled technologies, the QoS of V2X services is expected to be dramatically enhanced.
		\par
	\end{abstract}
	
	\section{Introduction}
	The anticipated development of autonomous driving and connected vehicles will bring numerous vehicle-to-everything (V2X) applications, including both safetyservice and entertainment  related services, which lead to a wide spectrum of Quality of Service (QoS) requirements in vehicular networks~\cite{R1}. Provisioning of V2X services require diverse data rate, reliability and latency from vehicular networks~\cite{3GPP-R2},~\cite{R3}. For instance, the autonomous driving requires a communication latency below a few milliseconds, and a reliability of close to 100\%. In comparison, entertainment services mainly focus on high data rate and thus can tolerate much longer latency and low reliability. Although Long Term Evolution (LTE) technology has incorporated V2X communication and its enhanced version (eV2X)~\cite{3GPP-R4}, current LTE networks still cannot support diverse QoS requirements of V2X services effectively with its ``one size fits all" architecture.
	\par
	Next generation wireless networks, i.e., fifth generation (5G) and beyond 5G, are expected to accommodate diverse V2X services. To achieve this goal, 5G wireless network is intend to adopt many emerging technologies. Apart from new radio access technologies (e.g., Non-Orthogonal Multiple Access (NOMA), massive Multiple-Input–Multiple-Output (MIMO)~\cite{R5}), network softwarization and intelligentization are the most distinctive and critical aspects of 5G networks for V2X service provisioning.
	\par
	Compared with existing rigid network architecture, network softwarization aims at providing highly flexible and programmable communication. Network softwarization a coherent integration of Software-Defined Networking (SDN) with Network Functions Virtualization (NFV). SDN can provide global view of network and programmable network control. On the other hand, through NFV, network functions and resources are not restricted to dedicated physical network infrastructures.
	Based on network softwarization, mobile network operator can allow service providers customize their own logical (virtual) networks, which can better guarantee the demand QoS of services, that is, “everything as a service” or “X as a service (XaaS)”~\cite{R61}. These logical networks are referred to as network slices,
	which are expected to utilize any available type of physical or virtual resource in wireless networks.
	Network slice
 	as a collection of Core Network (CN) and Radio Access Network (RAN) functions whose settings are configured to meet the diverse requirements of services \cite{R8,R10}. 
	Therefore, with network slicing, operator can flexibly compose slices for guaranteeing diverse requirements of different V2X services. Furthermore, it is also beneficial to implement the network slicing in vehicular networks, which can improve network agility and reconfigurability.
	\par
	On the other hand, precise perception of network environment and timely decision-making in 5G rely on intelligent operation due to highly complex and time-varying network conditions~\cite{R11}. With recent success of Artificial Intelligence (AI) enabled technologies, network intelligentization becomes a nature choice.
	Compared to conventional model-based system/network design methods, AI through Machine Learning (ML) methods, such as deep learning, can automatically extract dynamics of network and make decisions based on historical observations, without the need of expert and perfect knowledge of system~\cite{R12}. Furthermore, AI can be applied for different domains, ranging from network design, mobile data analysis, and network management. Therefore, it is necessary to utilize the AI-enabled technologies to facilitate deployment and management of wireless network, as well as realize automation and self-evolution of 5G network.
	\par
	Motivated by these two recent trends of 5G development, it is essential to utilize the network intelligentization in realizing the automated network slicing for V2X service provisioning. With AI-enabled technologies, we can intelligently extract high-level patterns of network slice that has complex structure and inner correlations. However, it is difficult to achieve through conventional model-based approaches. Therefore, the scope of this article is thus to bridge the gap between network softwarization and network intelligentization, by presenting the preliminary study on the ML based 5G network slice architecture for V2X services in vehicular networks. In the following sections, we aim to answer the key questions below in the intelligent network slicing, i.e.,
	\begin{enumerate}
		\item How can we design a flexible network slicing architecture, which can virtualize heterogeneous resources in vehicular networks for network slicing?
		\item How can we customize advanced ML methods for network slicing while considering specific requirement of vehicular networks?
	\end{enumerate}
	\par
	The remainder of this article is organized as follows. First, challenges in providing V2X services are discussed in Section II, where three typical V2X services and QoS requirements are summarized. Then, Section III presents the intelligent network slicing architecture for V2X services. Following that, Section IV discusses challenges and solutions for intelligent network slice architecture, which includes data collection and selection of deep learning method. Section V presents and discusses simulation results of previously introduced intelligent network slicing. Finally, conclusions are drawn in the last section.
	\section{Diverse Requirements of V2X Services}
	\begin{figure*}[!h]
		\centering
		\includegraphics[width=0.85\textwidth]{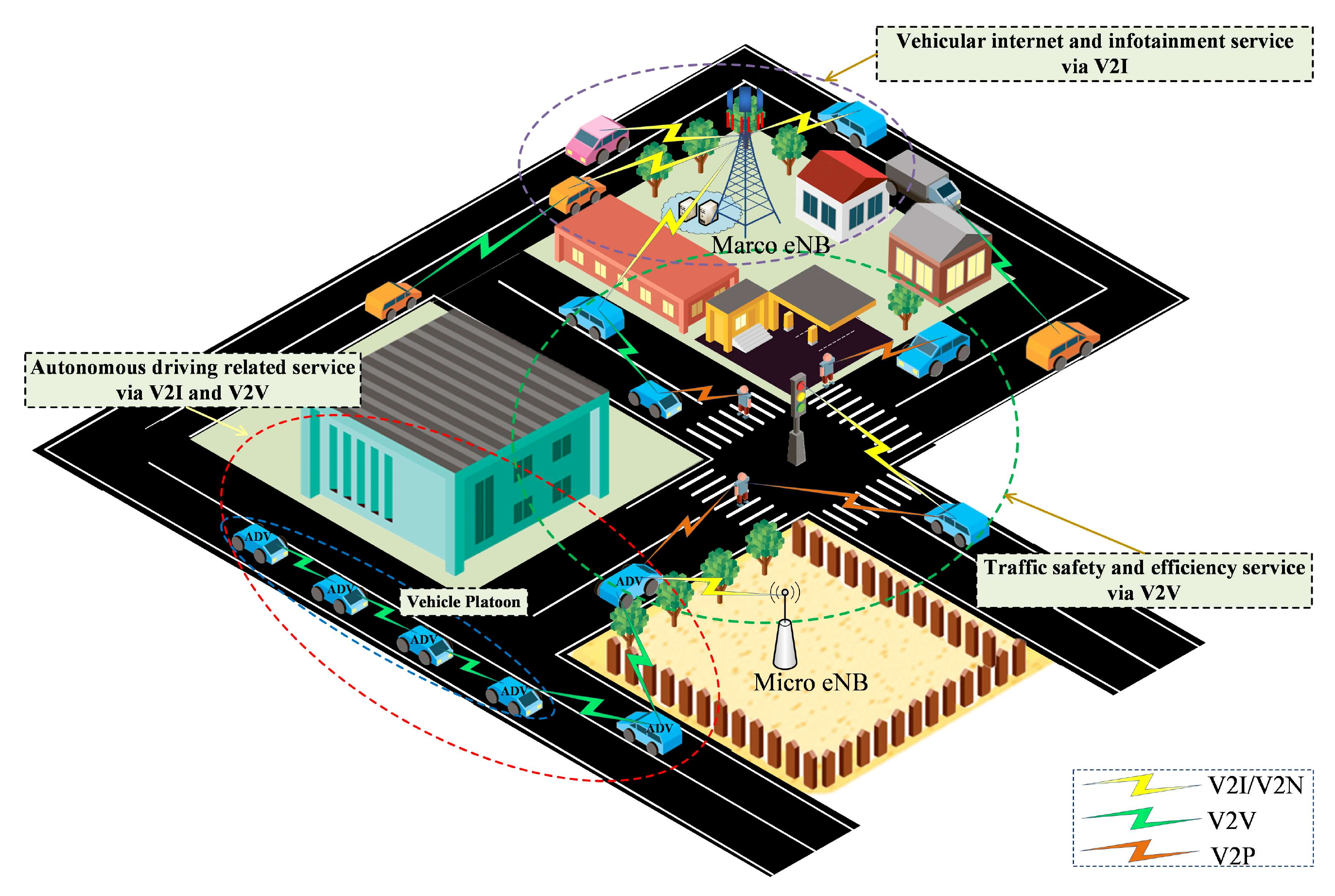}  
		\caption{Illustration of typical V2X services in vehicular networks.}
		\label{V2X services}
	\end{figure*}
To improve transportation efficiency, safety, and comfort on the road, V2X services have various and complex application scenarios, ranging from enhancing real-time safety and efficiency on road, to autonomous driving, to HD video streaming played for passengers in the vehicle.  To support V2X services with versatile communication type, V2X communication is proposed and classified into four patterns, i.e., Vehicle-to-Pedestrian (V2P), Vehicle-to-Vehicle (V2V), Vehicle-to-Infrastructure (V2I) and Vehicle-to-Network (V2N) communication. Based on the diverse service requirements, V2X services can be categorized into following three main types as shown in~Fig.~\ref{V2X services} and detailed requirements on QoS are summarized in Table 1. 
\begin{itemize}
	\item \textbf{Traffic safety and efficiency service}: It aims at reducing the possibility of traffic accidents and improvement of traffic efficiency. For instance, in order to avoid the accidents generated by blind spot, vehicles broadcast messages to surrounding vehicles via V2V communications. The periodic and event-driven messages carry the position and kinematics information of vehicle to allow other vehicles and pedestrians to sense the surrounding environment. Low latency and high reliability are considered as highly demanding requirements for this kind of services.
	\item \textbf{Vehicular internet and infotainment service}: The main objective here is to enable a more comfortable infotainment experience via V2N communication. Web browsing, social media access, and HD video streaming can provide on-demand entertainment information to drivers and passengers. It would become even more ordinary with more widespread of automobile driving, where drivers need media consumption.
	\item \textbf{Autonomous driving related service}: In principle, automated diverse is possible without V2X communication. However, the sensing ability of a single Autonomous Driving Vehicle (ADV) is limited. With the aid of cooperative awareness messages between other vehicles or infrastructures, the driving reliability of ADV can be significantly improved. For instance, ADV can also better adapt to the traffic situation by forming vehicle platoon and ADV can also download high-resolution digital maps by V2N communication. Compared to safety related service, ADV related service requires ultra-low latency, ultra-high reliability and high data rate requirements.
\end{itemize}
\begin{table}[!h]
	\centering
	\newcommand{\tabincell}[2]{\begin{tabular}{@{}#1@{}}#2\end{tabular}}
	\footnotesize
	\renewcommand{\arraystretch}{1.0}
	\caption{Typical V2X services and main QoS requirements.}
	\vspace*{-8pt}
	\label{table1}
	\begin{tabular}{|l|l|l|l|}
		\hline
		\textbf{Service category} & \tabincell{l}{Traffic safety and \\ efficiency service} &
		 \tabincell{l}{Autonomous driving \\ related service} &
		 \tabincell{l}{Vehicular internet and \\ infotainment service} \\
		\hline
		Communication Mode & V2V,V2P,V2I & V2V,V2I,V2N & V2N\\
		\hline
		Maximum Latency & 100 \textit{ms}~\cite{R3} & 10 \textit{ms}~\cite{R13} & \tabincell{l}{Low latency is not critical \\ for media streaming~\cite{3GPP-R2}}\\
		\hline
		Reliability Requirement & About 99\%~\cite{3GPP-R2} & 99.999\% \cite{R13} & Not a concern\\
		\hline
		Data Rate & About 1 Mb/s~\cite{3GPP-R2} & 10 Mb/s \cite{R13} & \tabincell{l}{0.5 Mb/s for web browsing, up \\ to 15 Mb/s for HD video \cite{3GPP-R2}}\\
		\hline
	\end{tabular}
\end{table}
	\par
	As mentioned before, network slicing can effectively cope with V2X services with divergent demands provided over 5G wireless networks. However, network slicing will introduce more complexity into already too complicated network, which make legacy network management routines untenable. Therefore, in the next section, we will elaborate the opportunity of utilizing AI for autonomous and effective network slicing.
	\section{Intelligent Network Slicing Architecture for V2X Services}
	In this section, we discuss how to design an intelligent network slicing architecture with high flexibility. Firstly, to provide a centralized controlling environment for network slicing, we present a cloud-based framework with the SDN technology for vehicular networks. Thanks to the centralization of the cloud-based framework, the intelligent network slicing architecture can conveniently be implemented for V2X services. The proposed architecture has two advantages. By using the NFV, different kinds of resource are virtualized from dedicated network infrastructures, and then network slice can be defined as a proper collection of resources, thereby further increasing flexibility. On the other hand, to efficiently manage network slice, AI-enabled technologies can be adopted while considering the characteristics of vehicular networks.
	\subsection{Cloud-based Framework for Vehicular Networks}
\begin{figure*}[!h]
		\centering
		\includegraphics[width=0.85\textwidth]{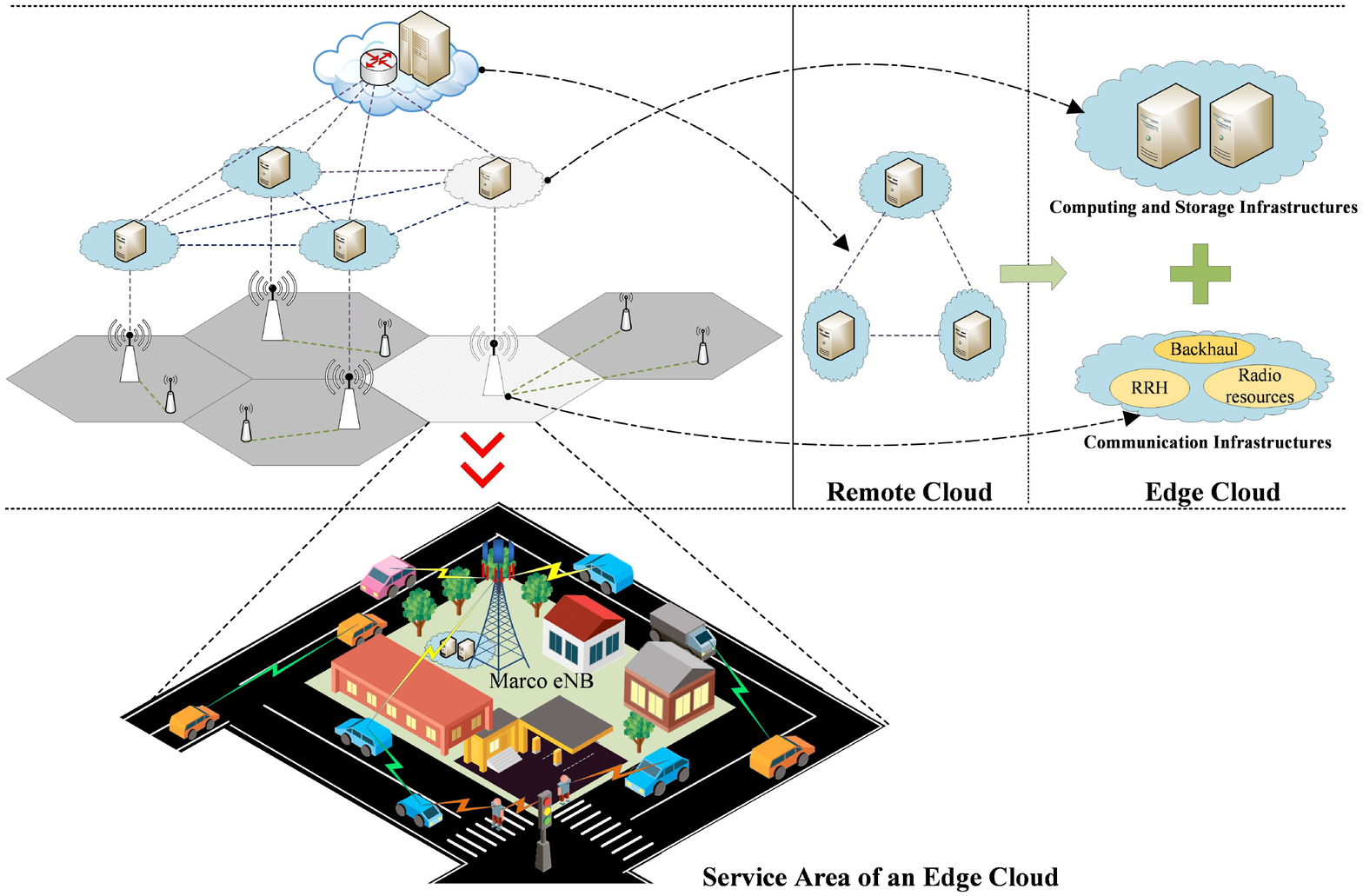}  
		\caption{Illustration of cloud-based framework for vehicular networks.}
		\label{cloud-based framework}
	\end{figure*}

	In order to perform a centralized control of network slicing, as shown in Fig.~\ref{cloud-based framework}, a cloud-based framework with SDN technology is presented by integrating various network infrastructures in vehicular networks. The cloud-based framework can be divided into two network domains, i.e., edge cloud and remote cloud.	
	\par
	
	\textbf{Edge Cloud}: The edge cloud, which hosts service entities in close physical proximity for Vehicular User Equipment (VUE), can ensure the low end-to-end response time of V2X services. The service area of edge cloud is defined as a basic geographical unit, which ranges from a single macrocell to a cluster of macrocells. When a VUE enters the coverage area of service area, it can enjoy V2X services host by the edge cloud. The edge cloud physically consists of communication, computation infrastructures and storage infrastructures. In the edge cloud, all network functions of RAN can be run as software modules on computation infrastructures. By jointly controlling communication, computing and storage infrastructures, edge cloud can provide V2X services to VUEs in a real-time way.
	\par
	\textbf{Remote Cloud}: For a certain edge cloud, there are surplus radio, computing and storage resources in other parts of vehicular networks. These resources can be defined as remote cloud. When the capability of edge cloud cannot meet the QoS requirements of V2X services, offloading the service requests of VUEs in its serivce area to the remote cloud becomes a good alternative. However, to access the remote cloud, it needs not only wireless links but also wired links, which will increase latency of V2X services.
	\subsection{Design of Intelligent Network Slicing Architecture}
	The centralized cloud-based framework provides the feasibility to apply the network slicing and ML based control for providing diverse V2X services. As shown in~Fig.~\ref{INS_Arch}, the intelligent network slicing architecture is proposed and divided as four layers: network infrastructure virtualization layer, intelligent control layer, network slice layer and service layer. The network infrastructure virtualization layer virtualizes different of resources from dedicated hardware of network without interoperability and compatibility issues, which makes network slicing more flexible and programmable~\cite{R8}. To deal with the increased complexity in network slicing, intelligent control layer can provide robust capabilities of orchestrating and managing of network slices.
	\subsubsection{Network infrastructure virtualization layer}
\begin{figure*}[!h]
	\centering
	\includegraphics[width=1\textwidth]{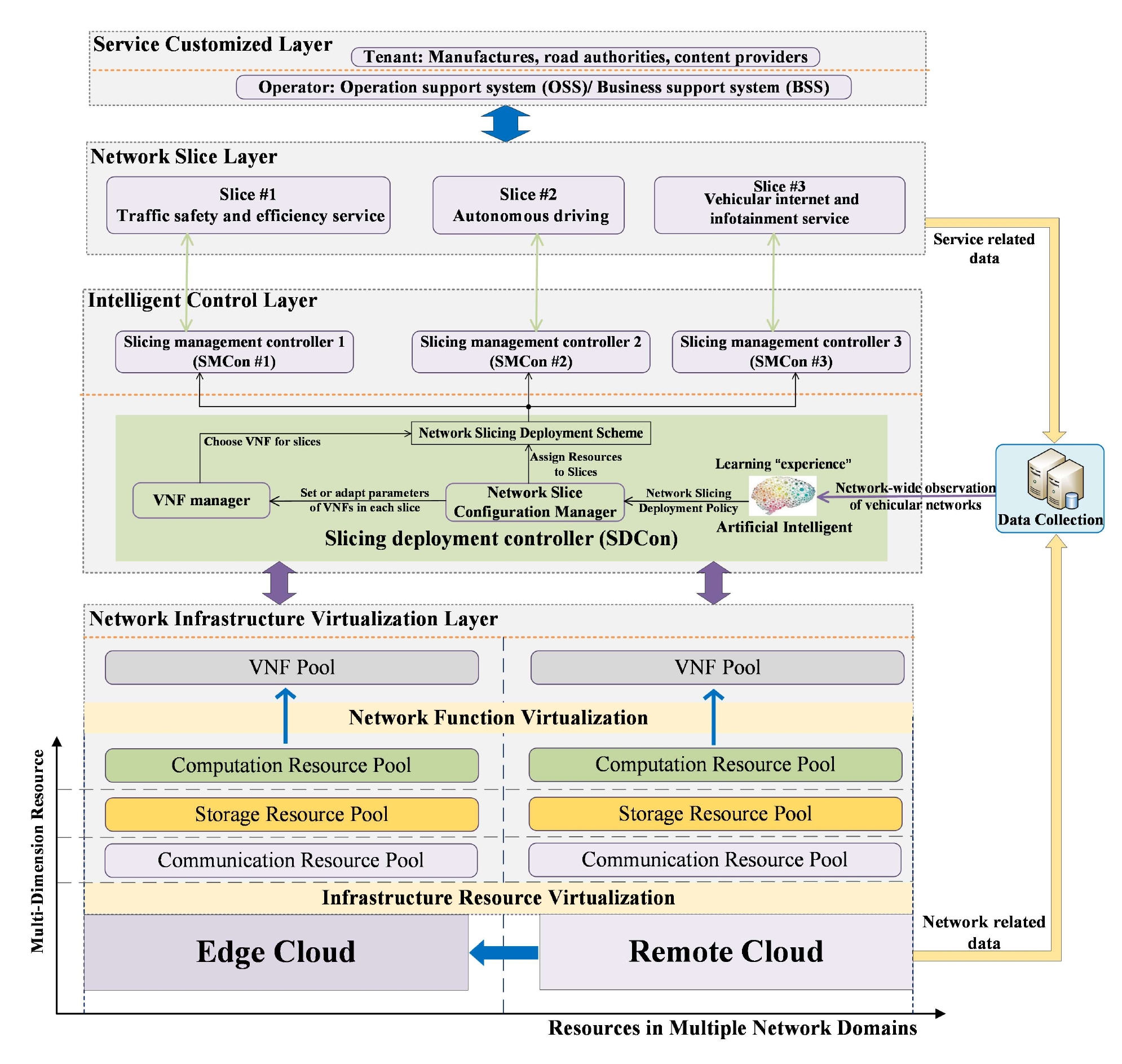}  
	\caption{Architecture of intelligent network slicing for V2X services.}
	\label{INS_Arch}
\end{figure*}

	As shown in the bottom of~Fig.~\ref{INS_Arch}, it is located at the bottom of the proposed architecture, which is responsible for (i) virtualizing, abstracting resources in underlaying infrastructures of each network domain (i.e., edge and remote cloud); (ii) decoupling network functions from dedicated physical hardware through NFV.
	\begin{itemize}
		\item \textbf{Infrastructure Resource Virtualization}: The resources in each network domain are mapped to the multi-dimension resource pool. According to its capability and ability, in the resource pool, resource is further divided into three dimensions: communication, computing and storage resource.
		\begin{itemize}
			\item \textit{Communication Resource Pool}: The communication resources consist of remote radio heads (RRHs), backhaul links and radio resources. Furthermore, radio resources can be further abstracted by a multi-dimensional grid of space, time, frequency and transmit power.
			\item \textit{Computation Resource Pool}: It has the virtualized information of all  computation resources in each network domain. Because computation infrastructures are heterogeneous, computation resource needs to be virtualized uniformly without concern about configuration details.
			\item \textit{Storage Resource Pool}: Similar to computation resources, abstraction of heterogeneous storage resources is also importance for unified management. Thus, storage resources in different infrastructures are integrated to the virtual storage pool to make them transparent to control layer.
		\end{itemize}
	\item \textbf{VNF Pool}: Virtual Network Function (VNF) can provide a variety of network abilities to support and realize required functionalities of V2X services. Generally, VNFs are implemented as software modules running with multi-dimension resources discussed above. A set of VNFs in different network domains are chosen to compose network slice. To make them available and programmable for control layer, we develop a VNF pool that integrates all individual VNFs distributed in each network domain. Hence, control layer can choose a set of VNFs in different network domains to compose a network slice as well as allocates network resources to support the operation of network slice.
	\end{itemize}
	\subsubsection{Intelligent control layer}
	Control layer is responsible for deploying and managing of network slices. Specifically, control layer exhibits operations by selecting appropriate VNFs for network slice, allocating and managing multi-dimension resources to host these VNFs. In addition, the proposed intelligent control layer also collects the mobile data generated by vehicular network, save it in database, and updated as well. Taking advantage of ML algorithms to analyze the mobile data, the intelligent control layer can extract pattern of vehicular networks and perform self-configuration of network slices to assure QoS requirements of V2X services.	
	\par
	However, due to the high mobility of vehicles, vehicular network topology is unstable and changing rapidly. Moreover, the size of vehicular network is normally very large. Therefore, ML algorithms normally need larger amount of mobile data and longer processing time to gain satisfying control performance. Consequently, if control layer directly utilizes ML algorithms to handle and optimize QoS performance of each V2X service request, the real-time V2X service is hard to guarantee due to the latency costed by signaling overhead and related processing complexity.
	\par
	
	Therefore, to perform deployment and management of network slices fast and efficiently, we design a hierarchical control layer, which can be further divided into two sub-layers: Slicing Deployment Controller (SDCon) and Slicing Management Controller (SMCon). From a two-timescale perspective, SDCon, an AI agent performs global control of network slicing, is responsible for deployment and configuration adaptation of network slices on a long-term timescale, while SMCon, a real-time regional control entity, directly manages multi-dimension resources in each network slice on a short-term timescale. Furthermore, SMCon, constituted by the VNFs with scheduling functionality, is defined and established by SDCon.
	\begin{itemize}
		\item \textbf{Slicing deployment controller (SDCon)}: SDCon lies in the bottom of the proposed control layer and has ability to deploy network slices according to customized V2X service requirements. By analyzing network-wide observations of vehicular networks (e.g., road condition and service traffic, etc.), SDCon, the AI learning agent, can achieve global status of vehicular networks, and adjust configurations of network slice to ensure the QoS requirements of V2X services. Furthermore, the network-wide observation data normally has obvious change on a long-term timescale (in level of second)~\cite{Mobile_Data}, SDCon should adapt configuration of network slices on a long-term timescale, which can also lower signal overhead and computation burden. SDCon decides network slicing deployment scheme, which includes,
		\begin{itemize}
			\item \textit{Orchestration of network slice}: A set of VNFs, combined with their interconnectivity, and the requested virtualized resources, which forms a deployed network slice. Firstly, in SDCon, VNF manager is responsible for mapping of VNFs in each network domain to the network slice and determining topology of the network slice. More particularly, in the VNF set, the VNFs used for scheduling intra-slice resources compose SMCon. Secondly, network slice configuration manager in SDCon is in charge of setting the necessary configuration for network slice instantiation and operation, which includes parameter settings of VNFs, allocating virtualized resources to host VNFs, and selection of V2X communication mode in the network slice.
			\item \textit{Network slice configuration adaption}: When current configurations of network slice cannot fulfill the desired QoS requirements of V2X service, network slice configuration manager will utilize ML algorithms to have an adjustment of slice configuration. Such adjustment may include modifications of specific VNF parameters and/or assigning more virtualized resources that dedicated to one network slice.
			\item \textit{Life cycle of network slice}: Manage life cycle of the network slices by setting trigger conditions and parameters to terminate network slice and release its resources.
		\end{itemize}
	\item \textbf{Slicing management controller (SMCon)}: Each SMCon, which is logically above SDCon, acts as a real-time entity to control its corresponding network slice directly. Based on the VNFs (e.g., radio resource scheduling and mobility management functions, etc.) determined by the network slicing deployment scheme, the SMCon is responsible for managing, allocating computation, storage and communication resources to VUEs with service request in its corresponding network slice. Furthermore, to quickly response on service requests from fast-moving VUEs, the SMCon should be realized by VNFs with low response latency.
	\end{itemize}
	\subsubsection{Network slice layer}
	After network virtualization and intelligentization of vehicular networks, network slices can be defined as an appropriate collection of multi-dimension resources in the multiple network domains. Then, based on the QoS requirements of typical V2X services stated in Table~\ref{table1}, we propose the following set of network slices.
	\begin{itemize}
		\item \textbf{Slice for traffic safety and efficiency service}: This slice supports the exchange of large amounts of message between many vehicles, SDCon should choose suitable resource scheduling function, which can avoid V2V channel congestion and ensure high reliable and low latency requirements.
		\item \textbf{Slice for autonomous driving}: To guarantee ultra-high reliable and ultra-low latency requirements of autonomous driving, SDCon should utilize semi-persistent scheduling function to avoid latency induced by signaling exchange, where adequate radio resources are reserved to ADVs.
		\item \textbf{Slice for vehicular internet and infotainment service}: To improve the overall user experience and reduce backhaul bandwidth use in V2N transmissions, SDCon should use caching functions to cache popular contents in advance.
	\end{itemize}
	\subsubsection{Service customized layer}
	The service customized layer is the top layer in the proposed architecture, which captures the QoS requirements of a given Service Level Agreement (SLA) as agreed by V2X service provider. Thus, mobile operator can support on-demand network slices for the V2X services.
	
	\section{Challenges and Solutions in Intelligent Network Slicing}
	From the perspective of network performance optimization, the realization of the proposed intelligent network slicing architecture in vehicular networks raises a number of related challenges. We mainly discuss the challenges of mobile data collection and implementation of deep reinforcement learning algorithm for SDCon in the intelligent control layer of the proposed architecture, in order to ensure the QoS requirements of V2X services and reduce the network cost for mobile network operators.
	\subsection{Mobile Data Collection for Intelligent Network Slicing}
	In the intelligent control layer, SDCon, the AI learning agent, focuses on understanding the dynamics of the global status of vehicular networks, and how to perform self-configuration and self-optimization of network slices to better accommodate it. Thus, collected mobile data, generated from heterogenous devices, needs to be pre-processed and converted into network-wide observation of vehicular networks, which can be used by SDCon to capture a global view on vehicular networks.
	\par
	\subsubsection{Mobile data in vehicular networks}

	In vehicular networks, both network side and user side are generating a huge amount of mobile data. Thus, we can collect two types of mobile data, namely service related data and network related data. The former is collected from VUEs in each network slice, while the latter one is obtained from the network infrastructure virtualization layer. Both types of mobile data are able to provide valuable insight into the global status of vehicular networks.
	\begin{itemize}
		\item \textbf{Service related data}: It refers to the data generated by VUEs in each network slice during the serving process. Service related data usually reflects VUE’s profile and communication pattern/behavior. Intelligent control layer should collect the location, velocity of vehicle, service request record, and QoS of applications. The collected data can be used by SDCon to characterize pattern of V2X services and improve the QoS of V2X services.
		\item \textbf{Network related data}: Mobile operators receive a large amount of status data generated by infrastructures in the edge and remote cloud. Then the network related information is assembled in the network infrastructure virtualization layer. Network related data can indicate network performance information and operating cost of vehicular networks. With network related data, such as multi-dimension resources status, interference between V2X links, network topology and energy consumption of infrastructures, SDCon can control the operating cost at a reasonable range.
	\end{itemize}
\begin{figure*}[!h]
	\centering
	\includegraphics[width=0.85\textwidth]{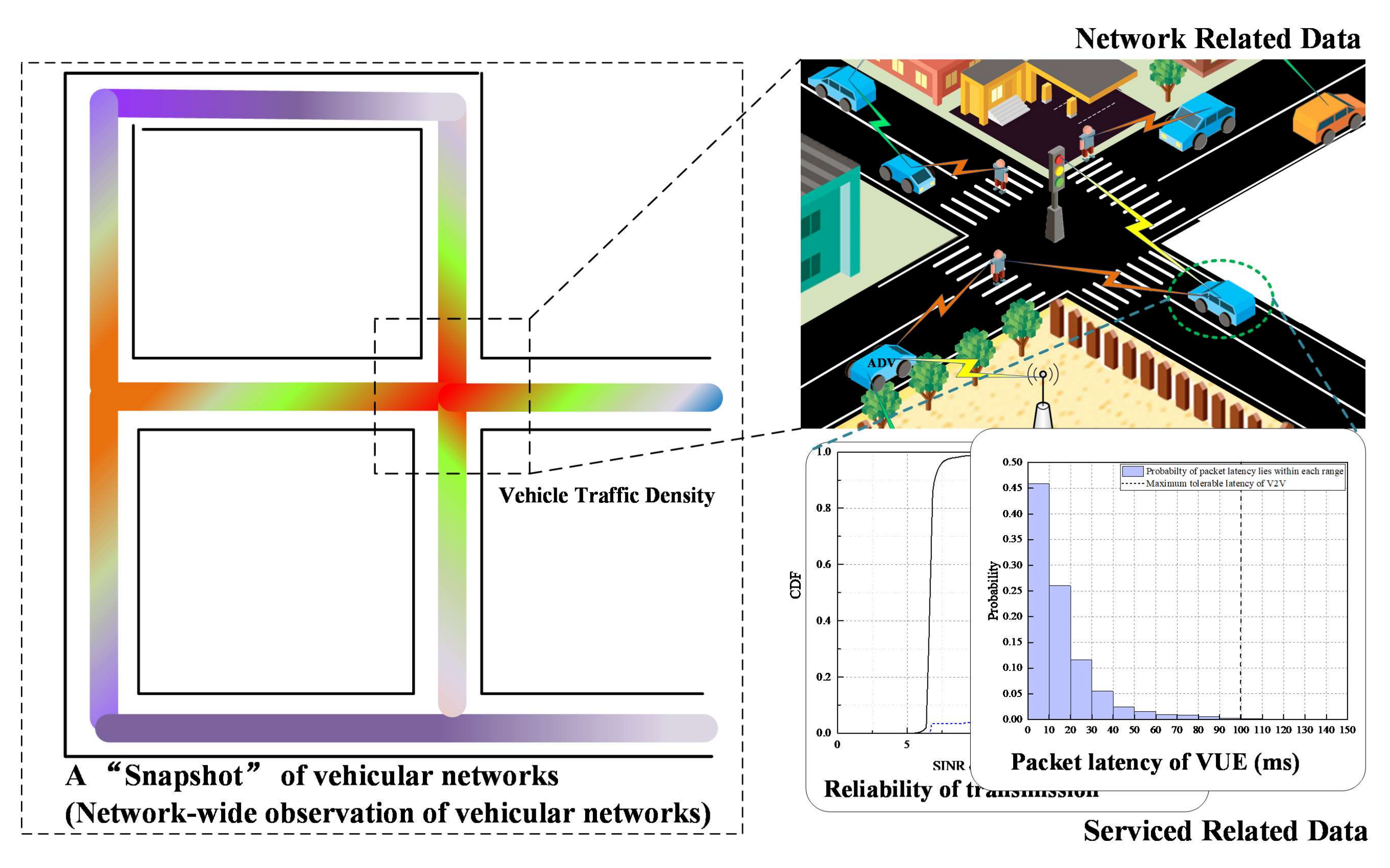}  
	\caption{Illustration of data collection and preprocessing. In a ``snapshot'' of vehicular networks, spatio-temporal dynamics of collected mobile data are aggregated over time. For instance, location and mobility of VUEs are converted to vehicle traffic density, and the areas marked in red has a higher vehicle traffic density than that marked in green or blue. Besides, QoS data of safety related service in this figure is cited from simulation result in our previous work~\cite{MJ}.}
	\label{data_collection}
\end{figure*}
	\subsubsection{Collection and preprocessing of mobile data}
	Due to the mobility of vehicles, mobile data in vehicular networks is exhibiting one significant feature: spatio-temporal diversity, i.e., from both spatial and temporal dimensions. Furthermore, it is recognized that mobile data in vehicular networks has important similarity with “videos”, which can be interpreted as a series of ``images''~\cite{R131}. According to this view, we can mark the collected mobile data on the map at its corresponding coordinate, which make a ``snapshot'', analogous to ``image'', of vehicular networks. A ``snapshot'' can be seen as a network-wide observation of vehicular networks. Then, in analogy with ``video'', a sequence of ``snapshots'' can be used to revel the dynamics of the global status of vehicular networks.
	\par
	On the other hand, since the ``snapshot'' (i.e., network-wide observation of vehicular networks) normally has obvious changes in level of second, it is not necessary for the intelligent control layer to collect mobile data in real-time. Instead, it is reasonable to collect mobile data on a long-term timescale to avoid the redundancy. Therefore, the intelligent control layer collects mobile data periodically and the period length of the data collection should be in level of second. Then, as shown in Fig.~\ref{data_collection}, the collection and preprocessing of mobile data can be described as follows:
	
	\begin{itemize}
		\item \textit{Step 1}: During the period, as shown on the right hand of Fig.~\ref{data_collection}, VUEs collect the service related data (e.g., location, mobility of vehicle, service request record, and QoS of applications) and infrastructures collect the network related data (e.g., resources status, V2X link quality, network topology and energy consumption of infrastructures). These raw data have the finest spatial and temporal granularity. However, it is inefficient to use these raw data to represent the network-wide observation of vehicular networks, which would induce data redundancy issues.
		\item \textit{Step 2}: At end of current period, VUEs and the infrastructures send its collected data to the intelligent control layer. Then, a ``snapshot'', as shown on the left hand of Fig.~\ref{data_collection}, is generated to represent network-wide observation of vehicular networks,
		\begin{itemize}
			\item The spatial granularity of the collected mobile data is decreased due to a limited number of collected data over large geographical regions;
			\item Temporal dynamics of the mobile collected data are aggregated over the period of time, and thus can better reflect the dynamics of the global status of vehicular networks.
		\end{itemize}	
	\end{itemize}
	\subsection{Intelligent Control Layer Powered by Deep Reinforcement Learning}
	In the intelligent controller layer, the key problem is how to let SDCon learn suitable deployment policy for network slicing, in order to provide satisfactory QoS for V2X services with satisfactory revenue. However, to achieve this goal, SDCon needs to deal with following two difficulties, i.e.,
	\begin{itemize}
		\item The QoS of V2X services is dependent on not only communication infrastructures but also resources and configurations of network slice.
		\item The resource demand of network slice is heavily related to the traffic pattern of V2X services. Thus, it is necessary to study how SDCon respond to traffic dynamics of V2X services in the long-term, in order to improve QoS of V2X service.
	\end{itemize}
	\par
	To settle them, the optimization of network slicing deployment scheme can be formulated as a Reinforcement Learning (RL) model, in which SDCon interacts with vehicular networks, then receives network operating revenue as feedback. The purpose of using RL is to find an optimal the deployment policy of network slicing deployment scheme to minimize network operating revenue. Furthermore, it is necessary to identify the actions, observations and reward functions in our DRL model, i.e.,
	\begin{itemize}
		\item \textit{Observation of vehicular networks}: The communication, computation, and storage status in each network domain, the QoS of service in each network slice, the vehicle traffic density in the service area, the number of VUEs in each network slice.
		\item \textit{Action taken by SDCon}: According to observations of vehicular networks, SDCon adjusts the network slicing deployment scheme by slice configuration adaption.
		\item \textit{Network operating revenue function}: It depends two factors. The first part is reward function in terms of latency and reliability requirements for V2X services. The second part is cost function for usage of communication, storage and computation resources.
	\end{itemize}
	\par
	However, in our RL model, the reward function (i.e., network operating revenue function) is related to large set of variables, which will bring the curse of dimensionality to traditional RL algorithms.
	To deal with the curse of dimensionality, Deep Reinforcement Learning (DRL) has attracted much attention. Due to Deep Neuro Network (DNN) can provide a good approximation of objective functions, researchers propose to further apply DNN to train a sufficiently accurate value/Q function and combine it with typical RL algorithms (e.g., Q-Learning). Generally, there are two kinds of DRL algorithm: value function methods (e.g., deep Q-learning) and policy search methods (e.g., policy gradient) through DNN. Value function methods are based on estimating the value function (expected return) of being in a given state. Compared to value function methods, the policy search methods do not need to maintain a value function model, but directly search for an optimal policy at cost of needing more training data samples and slower rate of convergence~ \cite{R15}.
	\par
	\begin{figure*}[!h]
		\centering
		\includegraphics[width=0.9\textwidth]{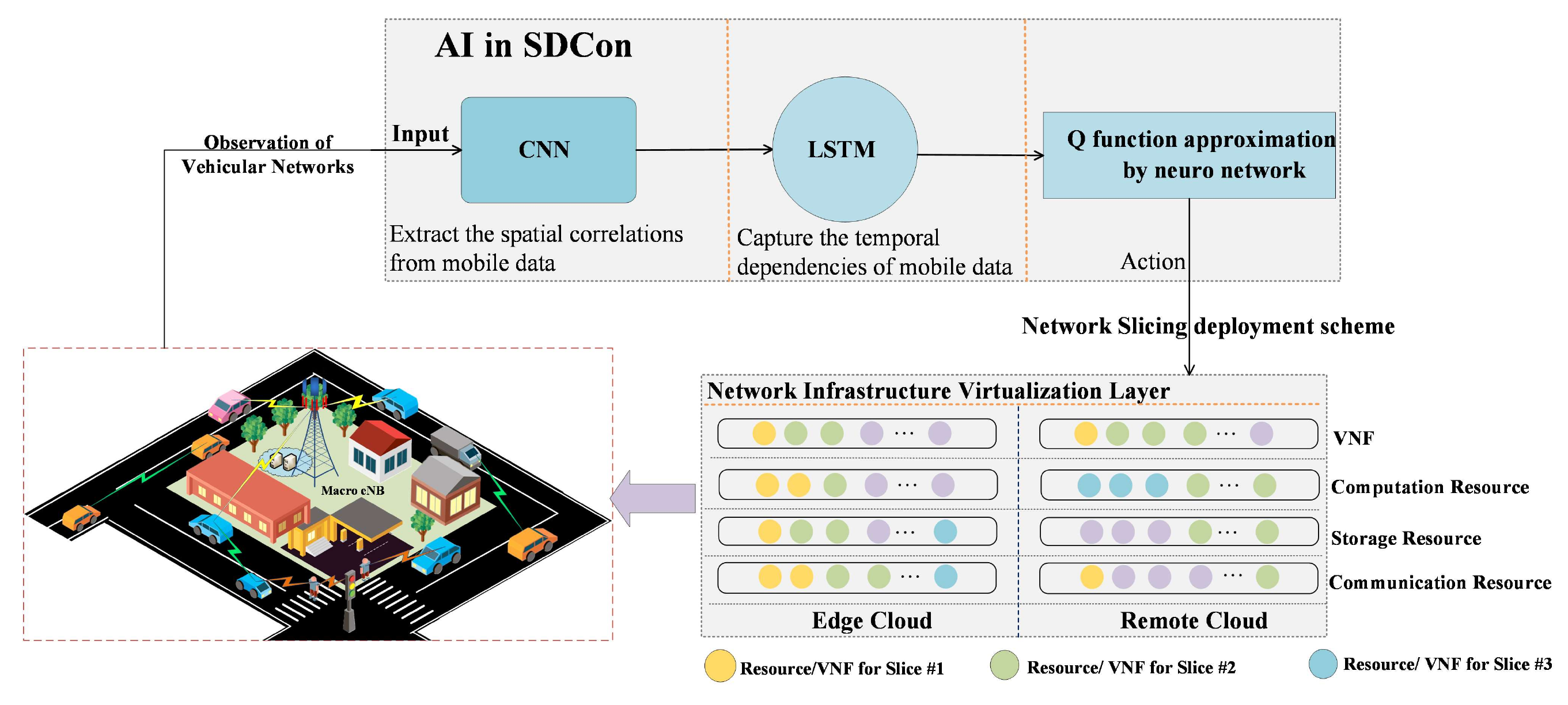}  
		\caption{An illustration of the structure of the proposed DRL algorithm used in SDCon.}
		\label{DRL}
	\end{figure*}
	We propose an improved deep Q-learning algorithm for minimizing the network operating revenue, where Deep Q-Network (DQN) map ``snapshot" (i.e., observation of vehicular networks) to the action taken by SDCon. As shown in Fig.~\ref{DRL}, two improvements are implemented in the proposed DQN,
	\begin{itemize}
		\item Utilizing Convolutional Neural Networks (CNN) to extract high-level features from raw input of the DQN (i.e., ``snapshot" of vehicular networks). The CNN employs a set of locally connected kernels filters to exploit the local spatial correlations of mobile data in different areas.
		\item Incorporating Long Short-Term Memory (LSTM) network into the DQN to maintain and memorize the observation history of vehicular networks, which can capture temporal dependencies between current observation and former states.
	\end{itemize}
	\par
	In general, combining both the above two techniques in the proposed DQN can better extract the spatio-temporal diversity of mobile data in vehicular networks. Thus, employing DRL algorithm at SDCon is promising to ``learn" a suitable network slicing deployment policy under complex, changeable, and heterogeneous vehicular network environments.
	\section{Simulation Results and Analysis}
	In order to demonstrate the effectiveness of our proposed intelligent slicing architecture, a system level simulation platform, an urban scenario based on the Manhattan grid layout defined in~\cite{3GPP-R2}, is implemented. Specifically, two types of V2V services (i.e., traffic safety and efficiency service, autonomous driving related service) are considered in the simulation platform.
	Furthermore, QoS of service is measured by two metrics, i.e., packet latency and Block Error Rate (BLER), which is used to characterize communication reliability.
	Detailed simulation parameters, including the system model and system assumptions, are summarized in Table~\ref{table2}. Software including  MATLAB 2017a and Keras 2.2.2 with Python 3.5.2. are utilized for simulations.
	\begin{table}[!h]
		\centering
		\newcommand{\tabincell}[2]{\begin{tabular}{@{}#1@{}}#2\end{tabular}}
		\scriptsize
		\renewcommand{\arraystretch}{1.0}
		\caption{Default Parameter settings for network slicing.}
		\vspace*{-8pt}
		\label{table2}
		\begin{tabular}{|l|l|l|}
			\hline
			\textbf{Parameter} & \multicolumn{2}{|l|}{\textbf{Assumption}} \\
			\hline
			Carrier frequency/Bandwidth & \multicolumn{2}{|l|}{2 GHz/10 MHz} \\
			\hline
			Number of eNB & \multicolumn{2}{|l|}{1}\\
			\hline
			Number of RBs/Bandwidth of each RB & \multicolumn{2}{|l|}{50/180 kHz} \\
			\hline
			Transmit Power of VUE & \multicolumn{2}{|l|}{6 dBm} \\
			\hline
			Pathloss model/Fast fading & \multicolumn{2}{|l|}{WINNER+ B1 Manhattan grid/Rayleigh fading}\\
			\hline
			VUE speed & \multicolumn{2}{|l|}{30 \textit{km}/\textit{h}} \\
			\hline
			& \multicolumn{2}{|l|}{Assign a set of RBs to each slice}   \\ 
			\cline{2-3}
			\tabincell{l}{Action taken by SDCon\\ (deployment scheme of network slices)} & \multicolumn{2}{|l|}{\tabincell{l}{Choose scheduling function for each slice: Round robin criteria, Channel\\ quality based criteria or Queue length based criteria}}   \\ 
			\hline
			Update cycle of slicing deployment & \multicolumn{2}{|l|}{100 $ms$ (i.e., 100 scheduling time units)} \\
			\hline
			V2V service type & Traffic safety and efficiency service & Autonomous driving related service \\
			\hline
			Packet Size & \tabincell{l}{Exponential distribution with mean \\ size 6400 bits} & Constant size 12800 bits \\
			\hline
			Arrival rate of packet & \tabincell{l}{Poisson distribution with average \\ arrival rate 0.02 packet/$ms$} & Message period is 10 $ms$ \\
			\hline
		\end{tabular}
	\end{table}
	\par

	Due to the simulation limitations, a simplified version of the DQN proposed in Fig. 5 is used in the simulation. It consists of input layer, LSTM layer, output layer: the input data of DQN are vehicle traffic density, service traffic in slices, and QoS of services; LSTM layer contains 128 units and uses Rectified Linear Unit (ReLU) as activation function; The hidden layer contains 24 units with linear activation function; The output layer gives evaluation of state-action value under a given input.
	To illustrate the importance of deep reinforcement learning (DRL) to the proposed architecture, we compare it with service-demand based network slicing method in~\cite{DRL_ref}: it assigns radio resources to each slice according to vehicle traffic density, latency and reliability requirements of V2X service. Under this method, round robin scheduling is conducted in each slice.
	\par
	\begin{figure*}[!h]
		\centering
		\includegraphics[width=0.85\textwidth]{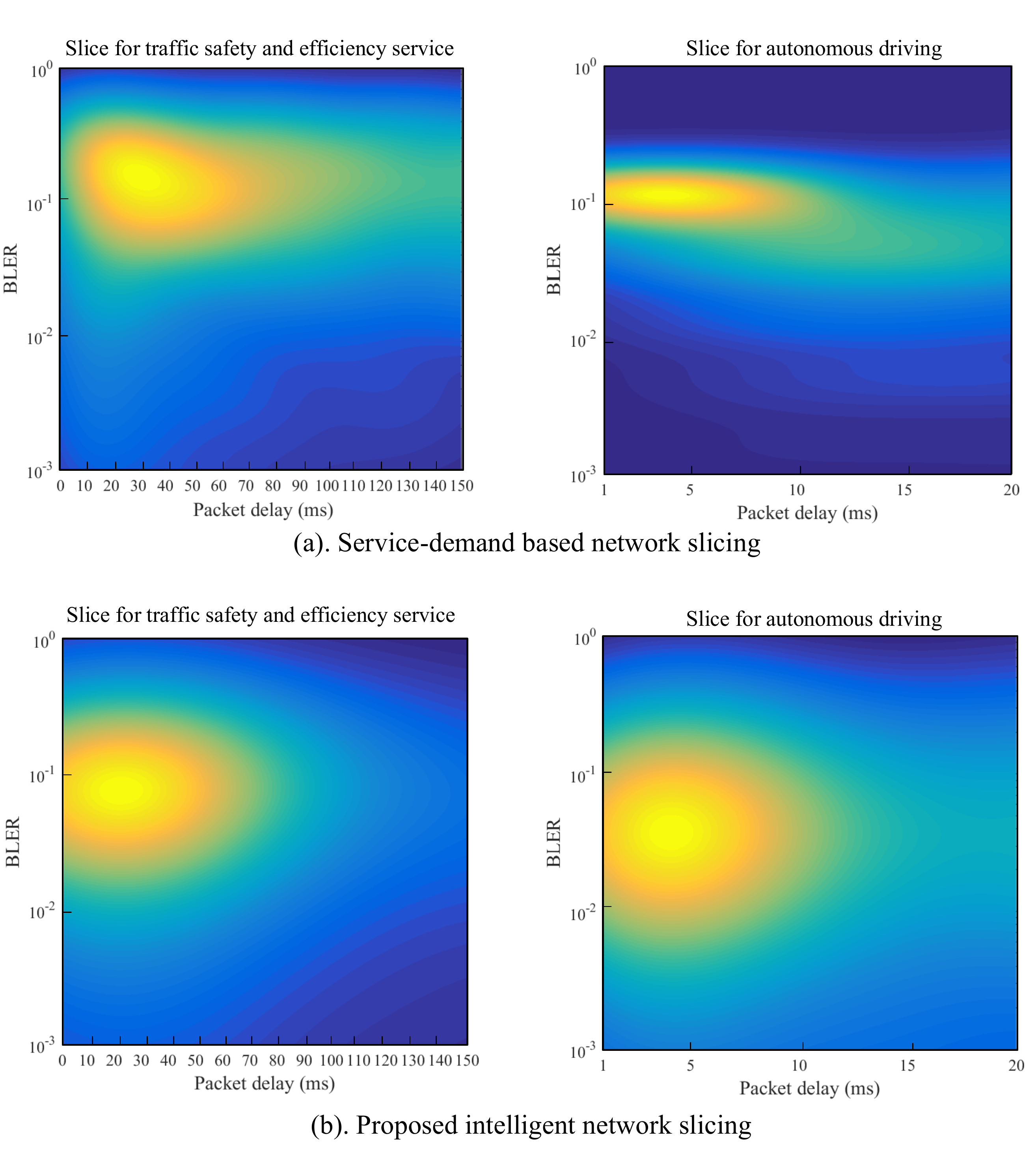}  
		\caption{Joint Probability Density Function (PDF) for BLER (i.e., reliability) and packet delay of V2V services in different network slicing methods: (a) Intelligent network slicing architecture. (b) Service-demand based network slicing. Besides, the area marked in yellow has a higher probability density than that marked in green or blue.}
		\label{Result}
	\end{figure*}
	By analyzing packet latency and BLER of each VUE from a statistical point of view, Figure~\ref{Result} presents the joint Probability Density Function (PDF) of BLER and packet latency, which reflects QoS of V2V service. Different colors represent different probability density values, i.e., the area marked in yellow has a higher probability density than that marked in green or blue. Under the proposed intelligent network slicing architecture, in Figure~\ref{Result}(b), it can be seen that there are more yellow colors (i.e., higher probability) in areas with lower BLER and lower packet latency than the reference network slicing method in Figure~\ref{Result}(a).  For traffic safety and efficiency service, packet latency is dramatically decreased. On the other hand, for autonomous driving related service, BLER is obviously decreased.
	The simulation results show that from both perspective of packet latency and BLER, the proposed network slicing architecture can improve QoS of V2V services, since it allows control layer to evolve the network slicing deployment policy according to history observations of vehicular networks.
	
	\section{Conclusion}
	Intelligent network slicing is a promising direction in meeting the diverse QoS requirements of V2X applications. 
	In this article, we have presented the design of intelligent network slicing architecture for V2X service provisioning. A new intelligent control layer, which is realized by a novel deep reinforcement learning algorithm, is proposed to achieve automated deployment of network slicing based on historical observations of vehicular networks. In addition, critical challenges of network slicing have been identified for its eventual deployment.
	Furthermore, we investigate the performance of our proposed intelligent network slicing by simulations to demonstrate its feasibility.
	However, this article does not consider issue of mining mobile data in vehicular networks, which can benefit multidisciplinary fields, ranging from transportation management and V2X services provision, to social network analysis and so on. On the other hand, due to limited radio resources, it is challenging to  provision sufficient orthogonal radio resources to each network slice. Therefore, many more works on intelligent network slicing are needed in order to fulfill the critical role of V2X in the anticipated autonomous driving era.
	\section*{Acknowledgment}
	This work is supported by the China Natural Science Funding under grant No. 61731004 and NSERC Discovery Grant RGPIN-2018-06254 of Canada. The authors would like to thank the anonymous reviewers for their constructive comments, which led to a significant improvement of this article.
	\bibliographystyle{IEEEtran}
	\bibliography{IEEEfull,INS_bibtex}
	\newpage
	\begin{IEEEbiography}
		{Jie Mei} (S'18) received his B.S. degree from Nanjing University of Posts and Telecommunications (NJUPT), China, in 2013. He received his Ph.D. degree in Information and Communication Engineering at Beijing University of Posts and Telecommunications (BUPT) in June 2019. Since August 2019, he is a Postdoctoral Associate with Electrical and Computer Engineering, Western University, Canada. His research interests include intelligent communications and Vehicle-to-Everything (V2X) communication.
	\end{IEEEbiography}

	\begin{IEEEbiography}
		{Xianbin Wang} (S'98-M'99-SM'06-F'17) is a Professor and Tier-1 Canada Research Chair at Western University, Canada. His research interests include 5G technologies, Internet-of-Things, communications security, machine learning and intelligent communications. Dr. Wang has over 380 publications, in addition to 29 patents and several standard contributions.
		\par
		Dr. Wang is a Fellow of Canadian Academy of Engineering, a Fellow of IEEE and an IEEE Distinguished Lecturer. He has served as Editor/Associate Editor/Guest Editor for more than 10 journals. Dr. Wang was involved in over 50 conferences with different roles such as symposium chair, tutorial instructor, track chair, session chair and TPC co-chair.
	\end{IEEEbiography}

	\begin{IEEEbiography}
		{Kan Zheng} (S'02-M'06-SM'09) is currently a full professor in Beijing University of Posts and Telecommunications (BUPT), China. He received the B.S., M.S. and Ph.D. degree from BUPT, China, in 1996, 2000 and 2005, respectively. He has rich experiences on the research and standardization of new emerging technologies. He is the author of more than 200 journal articles and conference papers in the field of vehicular networks, Internet-of-Things (IoT), security and so on. He holds editorial board positions for several journals and has organized several special issues. He has also served in the Organizing/TPC Committees for more than ten conferences such as IEEE PIMRC, IEEE SmartGrid and so on. 
	\end{IEEEbiography}
\end{document}